# Definition of scenarios for modern power systems with a high renewable energy share

**Authors:**

| |
|---|
| **Carlos Collados-Rodríguez** <br> CITCEA-UPC, Av Diagonal 647, H building, 2nd floor, Barcelona, Spain <br> ORCID: 0000-0002-5421-9775 |
| **Eduard Antolí-Gil** <br> CITCEA-UPC, Av Diagonal 647, H building, 2nd floor, Barcelona, Spain <br> ORCID: 0000-0002-8814-4468 |
| **Enric Sánchez-Sánchez** <br> CITCEA-UPC, Av Diagonal 647, H building, 2nd floor, Barcelona, Spain <br> ORCID: 0000-0003-4075-0191 |
| **Jaume Girona-Badia** <br> CITCEA-UPC, Av Diagonal 647, H building, 2nd floor, Barcelona, Spain <br> ORCID: 0000-0001-7842-6608 |
| **Vinicius Albernaz Lacerda** <br> CITCEA-UPC, Av Diagonal 647, H building, 2nd floor, Barcelona, Spain <br> ORCID: 0000-0001-8648-9027 |
| **Marc Cheah-Mañe** <br> CITCEA-UPC, Av Diagonal 647, H building, 2nd floor, Barcelona, Spain <br> ORCID: 0000-0002-0942-661X |
| **Eduardo Prieto-Araujo** <br> CITCEA-UPC, Av Diagonal 647, H building, 2nd floor, Barcelona, Spain <br> ORCID: 0000-0003-4349-5923 |
| ***Oriol Gomis-Bellmunt** <br> CITCEA-UPC, Av Diagonal 647, H building, 2nd floor, Barcelona, Spain <br> ORCID: 0000-0002-9507-8278 |

**Abstract**

Recent environmental policies have led many academic, industrial and governmental stakeholders to design and plan scenarios with very high share of renewable energy sources (RES). New system elements such as High-voltage Direct Current (HVDC) transmission, Microgrids, Virtual Power Plants (VPP) and Dynamic Virtual Power Plants (DVPP) are being increasingly studied with respect to their contribution to integrate future RES plants in the main grid. Several future scenarios are being analysed for each system and region, to ensure that the future energy systems, composed mostly of RES, can remain stable, match the demand during the seasonal variations across the year and are economically feasible. In this article, different types of energy scenarios are considered to obtain a range of options in terms of size, renewable generation technologies, and electrical network configuration. The scenarios were studied in the context of the POSYTYF project and were quantified through an optimization-based algorithm, using specific locations in Europe, and real data related to the availability of different RES, as well as the demand. It has been shown that photovoltaic (PV) and wind generation can provide the renewable backbone but they lack the flexibility needed to achieve a very high share in the energy mix. Other technologies, such as solar thermal and pumped hydro, become important to cover the last range of integration, as they provide a high flexibility, which is crucial for high share.

## 1. INTRODUCTION

The electrical power system is experiencing a deep transformation worldwide, due to the massive integration of renewable energy, the electrification of the demand and the irruption of electric mobility. This trend is intensifying and power systems have to achieve a massive share of renewable energy in next decades. Recently, several countries have defined targets to reduce the participation of fossil-fuel-based sources in the energy mix, while increasing the integration of Renewable Energy Sources (RES), such as wind, solar, geothermal, hydro, ocean and biomass. In the same direction, the European Union has set targets for specific levels of RES integration in the future European energy mix, with progressive participation of 20% in 2020 (European Council, 2009), 32% in 2030 and two thirds in 2050 (European Commission, 2012). These goals are to be achieved considering the participation of each Member State, which are defining its own policies and goals to match the general targets. For instance, Spain has established a target of 42% of RES share on energy end-use by 2030 (Spanish government, 2020).

Germany and France defined a target of 65% and 40% of RES in the final electricity consumption, respectively (European Commission, 2020).

In order to achieve the aforementioned targets, researchers need to use realistic power systems models considering a variety of technologies, system topologies and elements such as High-Voltage Direct Current (HVDC) (Van Hertem, Gomis-Bellmunt & Liang, 2016), Microgrids (Hatziargyriou, Asano, Iravani, & Marnay, 2007) , Virtual Power Plants (VPP) and Dynamic Virtual Power Plants (DVPP) (Marinescu, Gomis-Bellmunt, Dörfler, Schulte & Sigrist, 2021). Various future scenarios are being analysed for each system and region, to ensure that the future energy systems, composed mostly of RES, can remain stable, reliable, match the demand during the seasonal variations across the year and are economically feasible. These studies are regional by nature as they consider local weather and the availability of resources. Some examples were conducted for provinces or regions, such as Ontario (McPherson & Karney, 2017), British Columbia (Parkinson & Djilali, 2015), the New York State (Mahbub, Viesi & Crema, 2016), among others (Giallanza, Porretto, Puma & Marannano, 2018; Kalinci, 2015; Bracco, 2020; Ding, Liu, Huang Xu & Guo, 2019). Similar studies have also been performed using data from countries, such as Australia (Elliston, MacGill & Diesendorf, 2014), Bangladesh (Gulagi, Ram, Solomon, Khan & Breyer, 2020), Brazil (Schmidt, Cancella & Pereira, 2016; Dranka & Ferreira, 2018), Chile (Maximov, Harrison & Friedrich, 2019), France (Krakowski, Assoumou, Mazauric & Maïzi, 2016), Germany (Pregger, Nitsch & Naegler, 2013), India (Anandarajah & Gambhir, 2014), Italy, Pakistan (Sadiqa, Gulagi & Breyer, 2018), Portugal (Pina, Silva & Ferrão, 2013; Fernandes & Ferreira, 2014), United Arab Emirates (Almansoori & Betancourt-Torcat, 2015), the United States (Mai, Mulcahy, Hand & Baldwin, 2014). Other studies have also analysed systems with an ambitious goal of 100% of RES, such as (Luz & Moura, 2019; Henning, H & Palzer, 2014; Palzer & Henning, 2014; Norwood, Goop, & Odenberger, 2017; Zappa, Junginger & van den Broek, 2019). Moreover, as the number of studies has largely increased, several tools have been proposed to assist the generation expansion planning and RES design, such as EnergyPLAN (EnergyPlan, 2021), EnergyScopeTD (Limpens, Moret, Jeanmart & Maréchal, 2019), HOMER (HOMER Energy, 2021), LEAP (Stockholm Environment Institute, 2021), SILVER (McPherson & Karney, 2017), TIMES (IEA-ETSAP, 2021), among others (Connolly, Lund, Mathiesen & Leahy, 2010).

In this direction, the present paper summarizes several generation technologies and defines relevant future scenarios capturing the key features of the different renewable energy generation technologies, geographic and demand considerations and electrical topologies. The future scenarios were defined in the context of the POSYTYF project (POSYTYF, 2020). The presented concepts can be used as a start point to conduct more detailed other studies on different representative scenarios. Aspects related to cost, efficiency, resource availability and flexibility of different generation technologies are considered. Moreover, an optimization methodology is used to size the renewable power plants in different example scenarios, considering cost and availability. Therefore, this paper helps to understand the benefits of combining a wide range of different renewable energy generation technologies, where some provide generation at low cost but not controllable, while others provide more controllability at higher cost, but are fundamental for massive integration of renewables

The remainder of this paper is organized as follows. Section 2 briefly introduces each generation technology. Section 3 presents generated scenarios. Section 4 describes the methodology used to size the scenarios, including the optimization algorithm. Section 5 presents the defined scenarios resulting from the optimization algorithm. Finally, the conclusions are drawn in Section 6.

## 2. GENERATION TECHNOLOGIES

In this section, a brief review of the most relevant renewable and conventional generation technologies is presented, highlighting different characteristics that must be considered for an adequate sizing of the generation mix. These features are the following: response time, inherent storage time, controllability, dispatchability, $CO_2$ emissions and costs.

### 2.1 Solar photovoltaic

Photovoltaic (PV) systems encompass several PV modules (Figure 1). These modules are characterized by the well-known I-V curve which depends on external conditions like solar radiation levels and temperature. In order to obtain the maximum power output, the module must work as near as possible to the maximum power point (MPP) which is close to the knee of the I-V characteristic curve. For this purpose, power electronic devices such as inverters are constantly tracking the MPP considering solar radiation and temperature variations. Furthermore, these are employed for DC/AC conversion to connect the PV system into the grid. Although PV modules have negligible inherent storage capability, this can be provided by external devices.

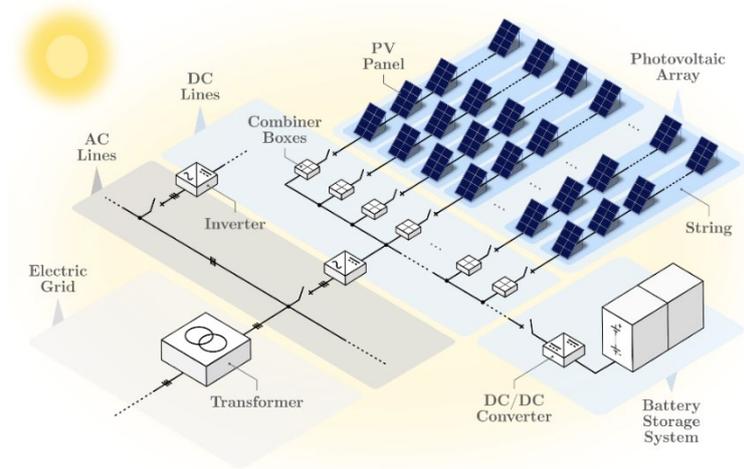

Figure 1. General scheme of a photovoltaic power plant.

**2.2 Solar thermal**

Solar thermal technologies use solar concentrators to produce the required high temperatures in the working fluid to raise steam to drive heat engines, mainly turbines in commercial plants. Therefore, solar concentrators perform a function similar to that of a boiler in a conventional thermal power plant based on a Rankine cycle. Steam temperature is critical to obtain acceptable conversion efficiencies. Nowadays, three proven technologies, which require direct or beam radiation, are appropriate for large-scale generation: parabolic troughs or linear Fresnel reflectors (both corresponding to linear focus technology), solar towers (Figure 2) or dishes (point focus technology). Depending on design details, large capacity thermal energy storage can be implemented, for instance, through molten salts.

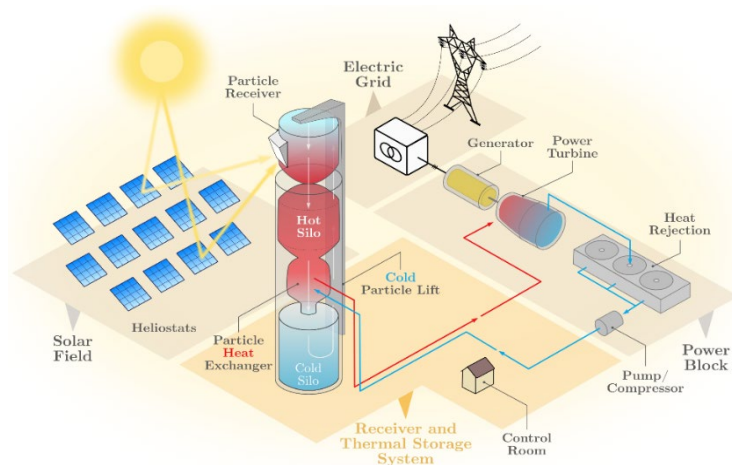

Figure 2. General scheme of a solar thermal power plant (solar tower).

**2.3 Wind**

Wind energy can be considered an indirect form of solar energy. Air flow is established due to the pressure gradient between high pressure and low-pressure zones, determining the initial speed and direction of wind flow.

Two types of wind farms can be distinguished: onshore wind farms (Figure 3) and offshore wind farms (Figure 4). Both types have several subsystems in common, such as AC connections between turbines, busbar, and transformer. Offshore wind farms might require exporting the generated power through HVDC technologies when these are located considerably far from shore (more than 80-100 km, approximately). Lastly, different grid topologies can be found based on its interconnection, e.g., radial, ring or star configurations (Van Hertem, Gomis-Bellmunt & Liang, 2016).

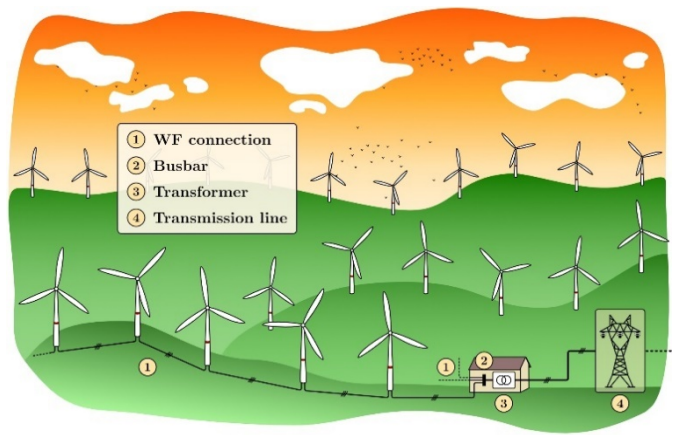

Figure 3. General scheme of an onshore wind farm.

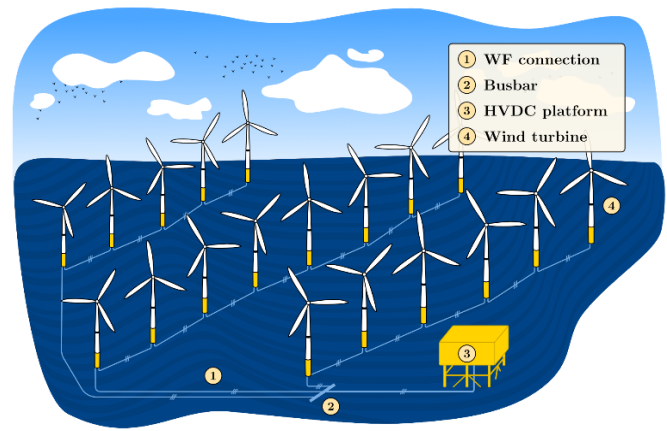

Figure 4. General scheme of an offshore wind farm.

**2.4 Hydroelectric**

Hydropower technologies take advantage of either water's potential or kinetic energy. Three main hydropower technologies can be distinguished: large scale hydropower (created by damming rivers), run-of-river hydropower and pumped-storage hydropower plants (PS-HPPs) (Figure 5). The suitability of each technology is highly dependent on the local topography (Infield & Freris, 2020).

In areas where the installation of large hydropower is unsuitable, PS-HPP is a promising alternative to consider. A PS-HPP comprises an upper and a lower reservoir and a binary or ternary pumping-turbine set, as shown in Figure 5. Whenever electricity is needed, water is driven from the upper reservoir to the lower reservoir and electricity is generated via the turbine system. When there is a surplus of electricity generation, water stored in the lower reservoir can be pumped back to the upper reservoir.

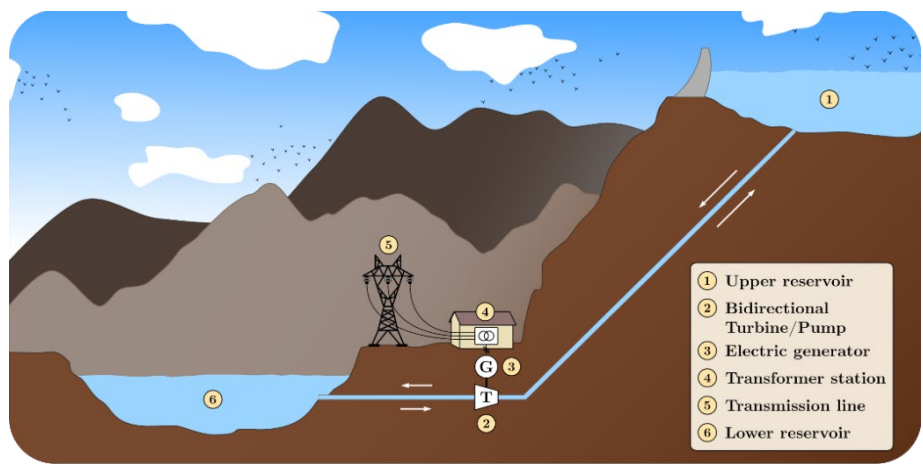

Figure 5. General scheme of a pumped-storage hydropower plant (PS-HPP).

**2.5 Biomass**

Biomass energy encompasses all sorts of solid biomass (such as wood, crops, etc.) or liquid biofuels that can be stored and used whenever required for electricity generation, similarly as fossil fuels, although with limited energy density. If it is possible, biomass must be produced and consumed locally (see right-hand side of Figure 6). That is the reason why most biomass power plants rely on local feedstock and supply chain. Besides, their size is usually smaller than conventional power plants. Regarding solid biomass, three thermochemical conversion technologies are distinguished: direct combustion, gasification and pyrolysis.

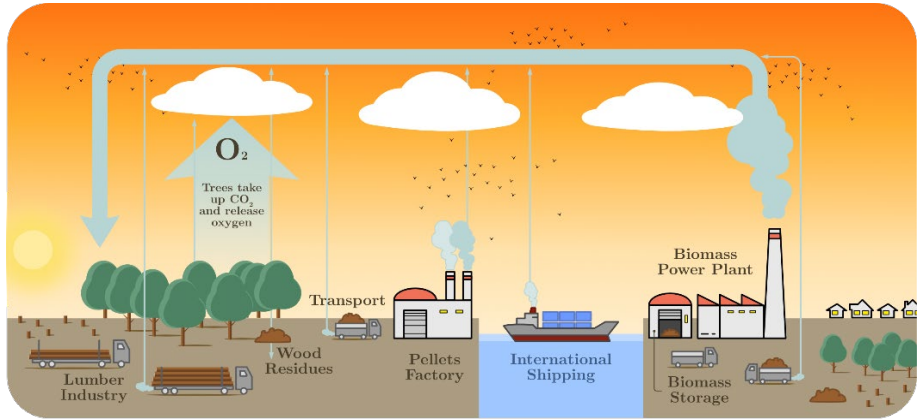

Figure 6. Conceptual scheme of the biomass resource process, including local generation and generation involving transport.

## 2.6 Geothermal

Geothermal energy derives from heat within the sub-surface of the earth (International Renewable Energy Agency (IRENA), 2020a). The heat transfer medium is water and/or steam. This renewable energy source is highly dependent on geographical locations. Besides electricity generation, if the temperatures are low, heat can be used for heating greenhouses, buildings or districts. Like other power plants, geothermal power plants use steam to drive steam turbines to produce electricity. A basic scheme of a generic geothermal power is shown in Figure 7.

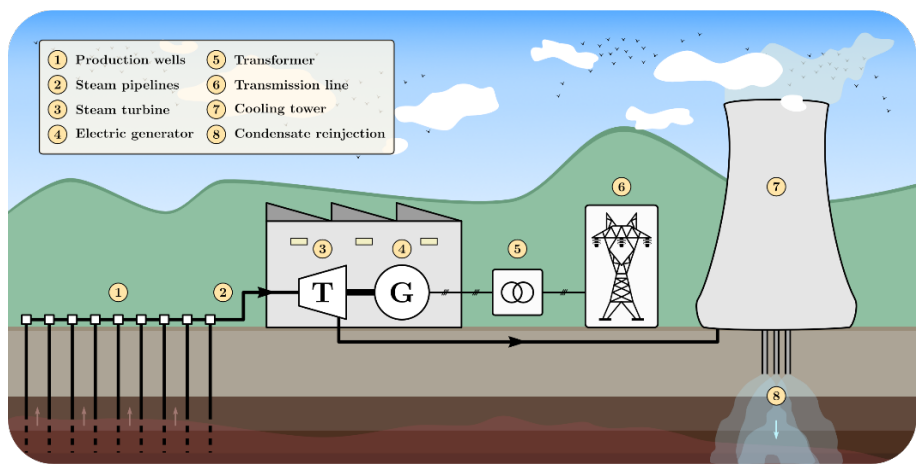

Figure 7. General scheme of a geothermal power plant.

## 2.7 Thermal coal/fuel

Conventional thermal power plants which use fossil fuels to generate electricity are based on a Rankine cycle (Figure 8). The coal/fuel burns inside the boiler, generating large amounts of heat used to produce highly pressurized steam. One or several sets of turbines (e.g., high, medium, or low pressure) generate rotating power via the aforementioned steam. Afterwards, the steam leaving the turbine's chamber is condensed using a cooling tower and, finally, recirculated back into the boiler to restart the cycle again (ENDESA, 2019a).

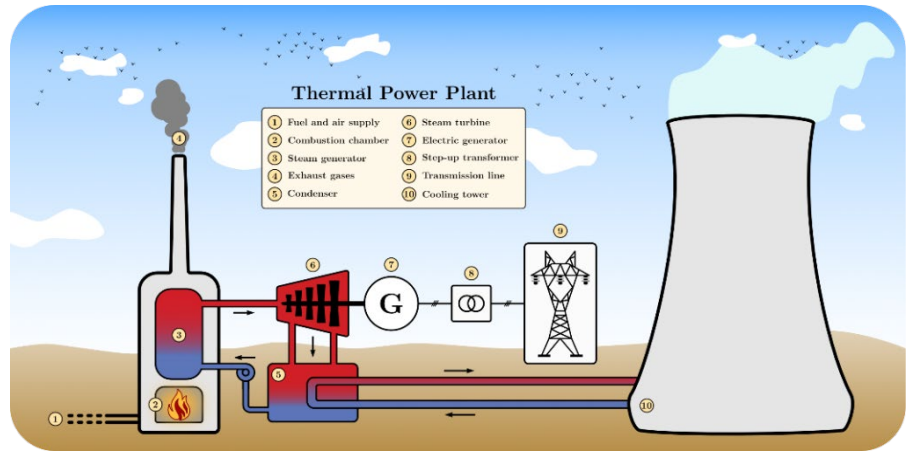

Figure 8. Conventional thermal power station.

**2.8 Thermal combined-cycle**

Combined-cycle power plants utilize natural gas to generate electricity (Figure 9). The plant bases its operation on two thermodynamic cycles: the Brayton cycle (gas turbine) and the Rankine cycle (steam turbine). Regarding the gas cycle, external air is compressed to high pressure through a compressor and mixed with gas. Then, the combustion takes place, and the combustion gasses expand in the turbine. Finally, the exhaust gasses are driven to a recovery boiler to raise steam for the steam cycle. Usually, both turbines are coupled to the same shaft (ENDESA, 2019b).

These power plants have higher efficiencies than conventional thermal power plants and can operate at a broader range of powers (min. 45% of the rated power). Moreover, their greenhouse gas emissions and refrigerating water consumption are lower. Also, for the same installed capacity, the infrastructure footprint is smaller.

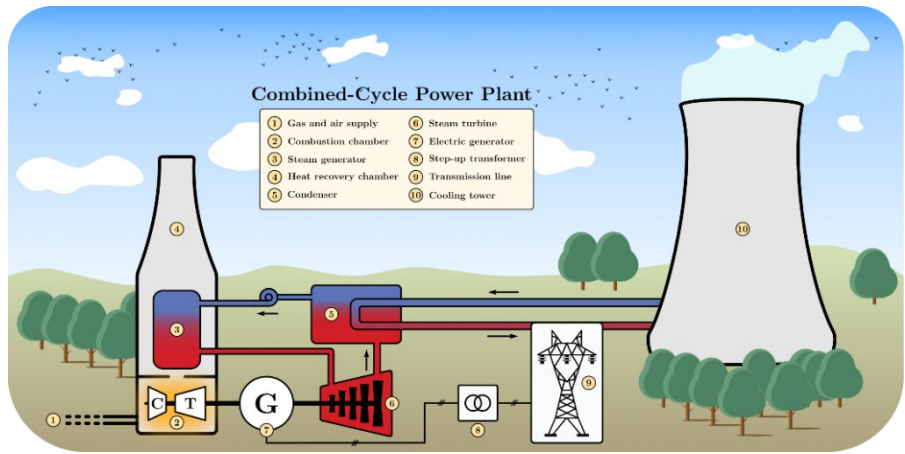

Figure 9. Combined-cycle thermal power station.

### 2.9 Nuclear

The most common reactor in a nuclear power plant is the Pressurized Water Reactor (PWR). Figure 10 sketches the main subsystems in a PWR nuclear power plant. Like thermal power plants utilizing fossil fuels, PWR plants are based on the Rankine cycle. However, in these power plants, heat is produced by fission in a reactor vessel containing water at very high pressure. Then, via a heat exchanger, the primary circuit transfers its energy to the secondary circuit.

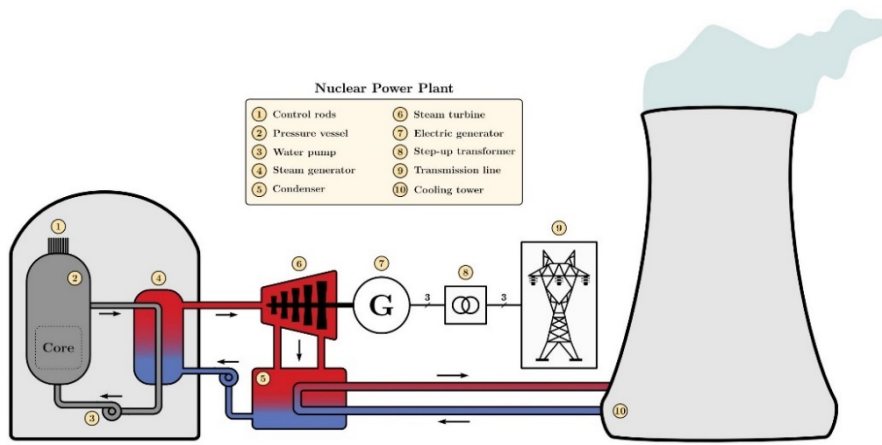

Figure 10. PWR nuclear power plant.

### 2.10 Metrics definition and summary

The following concepts are defined to compare the different features of each technology:

- **Controllability:** capability of a generation technology to store and control the power exchange with the network. Level definitions:

1. Non-storage capability. The resource defines the power injection to the grid. It can be only curtailed.

2. Limited storage of the converted energy. Example: thermal energy in solar thermal power plants can be stored.

3. Storage of primary energy - Low capacity

4. Storage of primary energy - Medium capacity

5. Storage of primary energy - High capacity

- **Dispatchability:** Capability of an electricity generation technology to provide power based on the operation setpoint (Tran & Smith, 2017). Level definitions:

  1. The primary energy availability permanently constrains the power output capability.

  2. The primary energy availability constrains the power output capability, but the power can exceed the threshold temporarily (short time-seconds)

  3. The primary energy availability influences the power output capability. However, the power output can be increased by means of a secondary (inherent storage) energy source.

  4. The primary energy availability is sufficient to not constrain the output power.

  5. The primary energy availability does not constrain the power output capability and it is possible to reverse the power plant to produce primary energy from the surplus of electricity in the network (bidirectional capability).

- **Response time:** The time elapsed between the acknowledgement of a new power reference and its successful tracking.

- **Inherent storage time:** Total amount of time that an electricity generation technology can provide electricity at full capacity by means of its inherent energy storage (Denholm & Mai, 2017).

- **$CO_2$ emissions:** Amount of $CO_2$ grams per kWh produced by an electricity generation technology considering its lifecycle footprint.

- **Levelized cost of electricity (LCOE):** Average revenue per unit of electricity generated that would be required to recover the costs of building and operating a generating plant during an assumed financial life and duty cycle (Energy Information Administration (EIA), 2020a).

- **Capital expenditure (CAPEX):** Funds used to acquire, upgrade, and maintain physical assets such as property, plants, buildings, technology, or equipment.

● **Operational expenditure (OPEX):** Expenses related to the production of goods and services.

Table 1 and Table 2 shows all previous characteristics for the different generation technologies. These aspects determine the role that each technology may have within the electric power system. PV and wind technologies present faster response times (from milliseconds to a few seconds) than other technologies solely based on synchronous generators. However, PV and wind inherent storage time are zero, whereas other technologies offer this characteristic, which ranges from hours to months (conventional plants).

Table 1. Technical characteristics of the different generation technologies considered.

|  | Response time | Inherent storage time | Controllability [1-5] | Dispatchability [1-5] | Generation technology |
|---|---|---|---|---|---|
| **PV** | 100 ms - 5 s [6] | 0 | 1 | 1 | PE |
| **ST** | 15 min – 4 h [a][7] | 0 - 24 hours [8] | 2 | 3 | SG |
| **W** | 0.5 ms - 1 s [9] | 0 | 1 | 2 | SG/IG+PE |
| **HYD** | 2 - 5 min [10] | 4h - 16h [11] | 3 | 4 | SG |
| **BIO** | 10 min – 6 h [b][7] | Weeks | 4 | 4 | SG |
| **CF-TPS** | 80 min - 8 h [12] | Months | 5 | 4 | SG |
| **CC-TPS** | 5 min – 3 h [12] | Months | 5 | 4 | SG |
| **N-TPS** | ~24 h [7] | 18-24 Months | 5 | 4 | SG |
| **PS-HPP** | 2 - 5 min [10] | 4h - 16h [11] | 3 | 5 | SG |
| **GEO** | 30 s – 2 min | inf | 5 | 4 | SG |

[a] Ramping rate: 6% of full load/min. Hot start-up time: 2.5 h
[b] Ramping rate: 8% of full load/min. Hot start-up time: 3 h

[1] (EUROPEAN COUNCIL. 2009), [2] (EUROPEAN COMMISSION, 2012), [3] (SPANISH GOVERNMENT, 2020), [4] (ENDESA, 2019A), [5] (MCPHERSON & KARNEY, 2017), [6] (BULLICH. MASSAGUÉ, FERRER.SAN.JOSÉ, ARAGÜÉS.PEÑALBA, SERRANO.SALAMANCA, PACHECO.NAVAS & GOMIS.BELLMUNT, 2016), [7] (GONZALEZ-SALAZAR, KIRSTEN & PRCHLIK, 2018), [8] (OFFICE OF ENERGY EFFICIENCY & RENEWABLE ENERGY, 2020), [9] (ABU-RUB, MALINOWSKI & AL-HADDAD, 2014), [10] (EUROPEAN ENERGY RESEARCH ALLIANCE (EERA), 2016), [11] (ENVIRONMENTAL AND ENERGY STUDY INSTITUTE (EESI), 2019), [12] (INTERNATIONAL RENEWABLE ENERGY AGENCY (IRENA), 2018)

Legend: PV: solar photovoltaic – ST: solar thermal – W: wind – HYD: hydropower – BIO: biomass – CF-TPS: coal-fired thermal power station – CC-TPS: combined-cycle thermal power station – N-TPS: nuclear thermal power station – PS-HPP: pumped-storage hydropower plant – GEO: geothermal – PE: power electronics – SG: synchronous generator – IG: induction generator.

Table 2. Costs and emissions of the different generation technologies considered.

|  | LCOE [$/kWh] | CAPEX [1] [$/kW] | OPEX [1] [$/kW] | Fuel cost | $CO_2$ Emissions [2][3] [g-eq/kWh] |
|---|---|---|---|---|---|
| **PV** | 0.029 to 0.042 [4] | 1313 | 15.25 | 0 | 18 to 180 |
| ST | 0.126-0.156 [4] | 7221 | 85.40 | 0 | 9 to 63 |
| W | 0.026 to 0.054 (onshore), 0.086 (offshore) [4] | 1265 to 4375 | 26.34 to 110 | 0 | 8 to 40 |
| HYD | 0.0473 [5] | 5316 | 29.86 | 0 | 2 to 200 |
| BIO | 0.0656 [5] | 4097 | 27.47 | 20-50% LCOE | 50 to 400 |
| CF-TPS | 0.065 to 0.159 [4] | 3676 to 5876 | 40.58 to 59.54 | 42.47 $/t [6] | 850 to 1125 |
| CC-TPS | 0.044 to 0.073 [4] | 958 to 2481 | 12.20 to 27.60 | 0.106 $/m$^3$ [7] | 450 to 525 |
| N-TPS | 0.129 to 0.198 [4] | 6041 to 6191 | 95.00 to 125.72 | 3-5 €/MWh [8] | 15 to 30 |
| PS-HPP | 0.0473 [5] | 5316 | 29.86 | 0 | 2 to 200 |
| GEO | 0.059 to 0.101 [4] | 2521 | 129.70 | 0 | 50 |

[1] (ENERGY INFORMATION ADMINISTRATION (EIA), 2020B), [2] (THE INTERGOVERNMENTAL PANEL ON CLIMATE CHANGE (IPCC), 2014), [3] (INTERNATIONAL INSTITUTE FOR APPLIED SYSTEMS ANALYSIS (IIASA), 2012), [4] (LAZARD, 2020), [5] (INTERNATIONAL RENEWABLE ENERGY AGENCY (IRENA), 2020B), [6] (ENERGY INFORMATION ADMINISTRATION (EIA), 2020C), [7] (ENERGY INFORMATION ADMINISTRATION (EIA), 2020D), [8] (RICO, 2014)

## 3. SCENARIOS GENERATION

In this section, several scenarios are defined in order to illustrate different power systems with different characteristics, such as the grid configuration or the combination of RES technologies in the system. The classification criteria are defined as follows:

- Three main grid configurations:
    - Type I: Isolated
    - Type II: Synchronously interconnected (AC)
    - Type III: Non-synchronously interconnected (DC) (i.e. isolated systems with only DC interconnection/s)
- Combination of different RES technologies:
    - Different portion of RES in the system
    - Controllable and non-controllable technologies
    - Consider power electronics in the generation plants
- In terms of grid layout, only transmission, or transmission plus distribution
- Optionally, non-electrochemical storage can be included

Based on the already existing scenarios in Europe, four realistic scenarios have been built as examples of power systems based on the different previously mentioned characteristics:

- Type I: island scenarios are, in general, smaller and more straightforward than continental ones. Therefore, a smaller number of buses (in this case, seven) and a single voltage level is considered for this case (Figure 11).

- Type II: the majority of scenarios are AC interconnected systems, and they are typically bigger and highly meshed. Consequently, a higher number of buses (in this case, thirteen) and different voltage levels (i.e. transmission and distribution) are considered. Moreover, two distinct versions of this type of scenario are considered. One corresponds to a typical southern Europe scenario (Figure 12), whereas the other corresponds to a typical northern Europe scenario (Figure 13), including HVDC interconnected offshore wind.
- Type III: regarding HVDC interconnected scenarios without AC interconnections, they typically correspond to bigger islands. For that reason, the grid layout considered is slightly more complex, with a higher number of buses than Type I. Additionally, different voltage levels are considered in this case (Figure 14).

It should be noted that these scenarios are still preliminary since the power ratings of the transmission lines and the generation units are not defined. Based on these layouts, the algorithm described in Section 4 assigns an optimal rating to each element in the system, considering several inputs and restrictions.

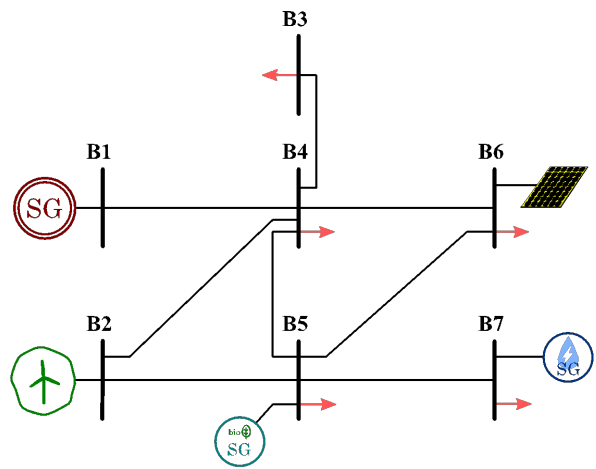

Figure 11. Selected scenario 1: type I.

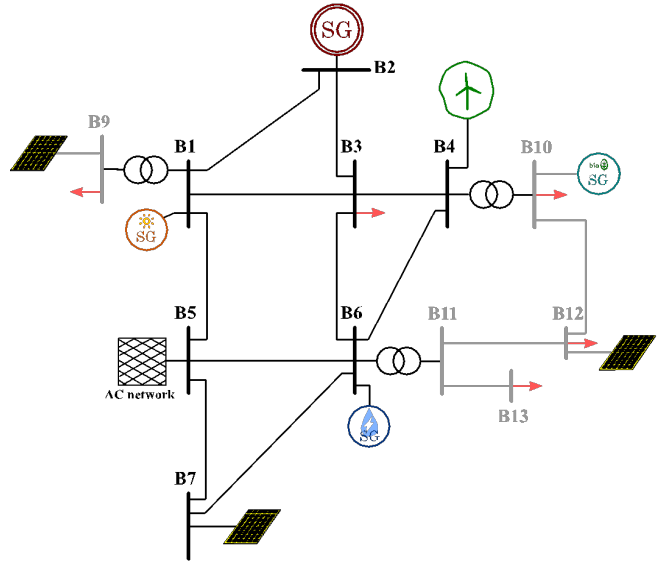

Figure 12. Selected scenario 2: AC interconnected (type II, southern Europe).

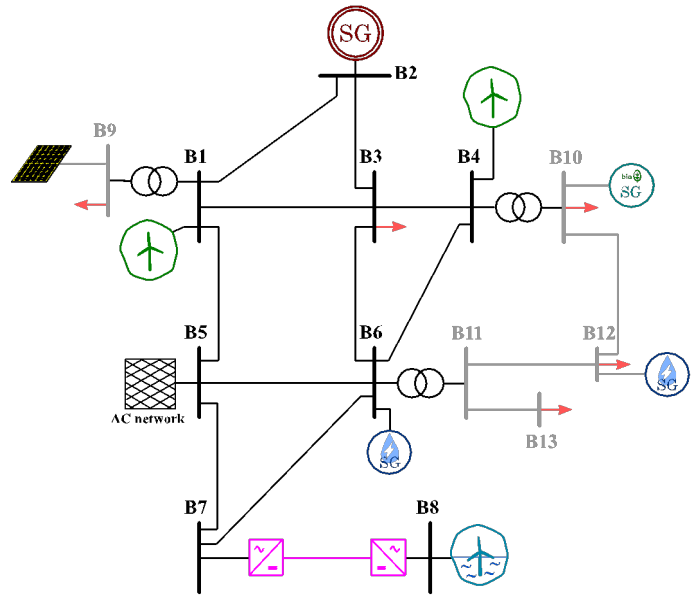

Figure 13. Selected scenario 3: AC interconnected (type II, northern Europe)

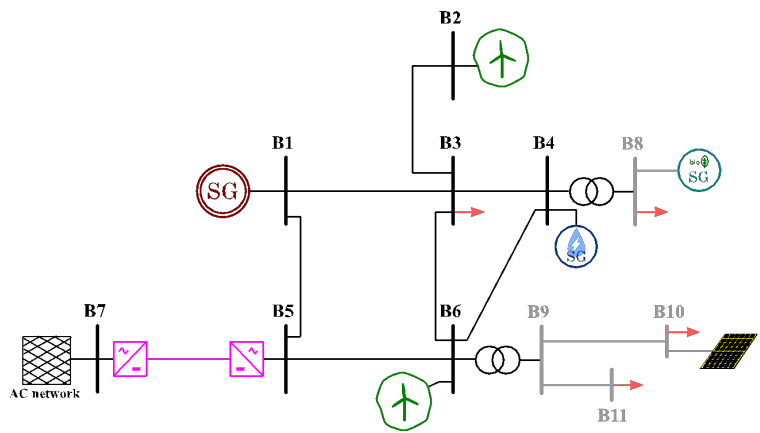

Figure 14. Selected scenario 4: DC interconnected (type III)

# 4. METHODOLOGY FOR SIZING THE SCENARIOS

In this section, a methodology to size the renewable generation technologies for realistic scenarios is described. The objective is to quantify the elements that are included in the previous scenarios, such as the rated capacity of the power plants or the capacity and length of the transmission lines. The sizing methodology is based on a generation cost optimization, considering the European or the local policies regarding the objectives of renewable generation. The grid restrictions are not considered in this algorithm. The optimization quantifies the renewable generation that should be installed to fulfil the minimum share of renewable generation while minimizing the total generation cost.

Based on the characteristics and system topologies mentioned above, a large number of scenarios can be generated. Then, the optimization algorithm has been applied only to the Scenario 1 in Figure 11 and Scenario 3 in Figure 13 to exemplify potential results that could be obtained using this methodology. New scenarios can be easily generated by modifying the initial ones.

## 4.1 Generation cost optimization

The optimization algorithm has been developed in Python in order to obtain the renewable capacity that minimizes the generation costs. A flowchart of this algorithm is shown in Figure 15. Several inputs are required in order to define the power plants and system characteristics:

- Conventional generation: it is assumed to be already installed in the system, so CAPEX is not considered. Then, the inputs required for the conventional thermal power plants are the installed capacity and the OPEX.
- Renewable generation: both CAPEX and OPEX are considered as inputs. In addition, the availability of resources, *i.e.* irradiation or wind speed, is also required. If a renewable power plant has already been built, the CAPEX is no longer needed, but the installed capacity is instead.
- System: the total demand at each time interval and the minimum share of renewable generation.

The previous inputs are specifically defined for each conventional and renewable generation model.

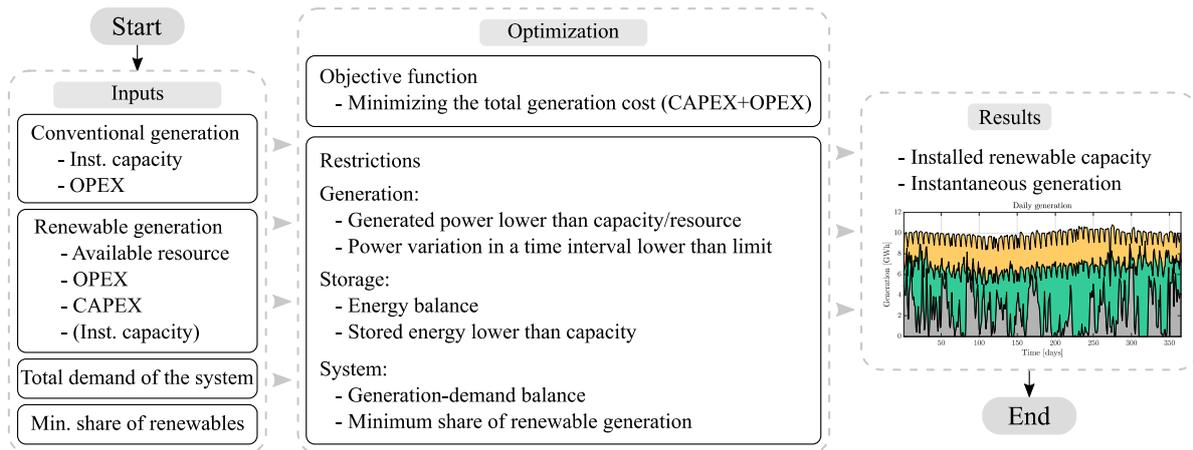

Figure 15. Overview of the optimization algorithm

## 4.2 Modeling of the system elements

Models for conventional and renewable generation have been implemented in Python to represent the particular characteristics of every energy resource. Four models have been considered: conventional power plants, renewable power plants without storage (PV and wind), solar thermal power plants and pumped-storage hydropower plants. System restrictions are also included in the model.

*4.2.1 Conventional power plants*

Conventional power plants, *e.g.* coal or gas-based power plants, are represented by the following restriction:

1. Maximum power generation: the instantaneous power generation must be lower or equal to the installed capacity of the power plant.

$$x_{GCi,t} \leq G_{Ci} \quad \forall i, \forall t \qquad (1)$$

where $i$ denotes each of the $C_i$ conventional power plants, $x_{GCi,t}$ is the instantaneous generation of the conventional power plant $C_i$ at the time $t$ and $G_{Ci}$ is the installed capacity of the conventional power plant $C_i$.

*4.2.2 Renewable generation without storage (PV and wind)*

The modeling of PV and wind power plants is similar to that in conventional power plants, as they present the same restriction. However, in this case, the maximum generation will depend on the availability of the resource:

1. Maximum power generation:

$$x_{GRj,t} \leq G_{Rj} \cdot C_{Rj} \quad \forall j, \forall t \quad (2)$$

where $j$ refers to each of the $R_j$ renewable power plants, $x_{GRj,t}$ is the instantaneous generation of the renewable power plant $R_j$ at the time $t$, $G_{Rj}$ is the installed capacity of the renewable power plant $R_j$ and $C_{Rj}$ is the available resource expressed in per unit. The solar and wind resources can be obtained for a specific location and time interval in (Renewables.ninja, 2021).

*4.2.3 Pumped-storage hydropower plants*

The hydropower plants have been considered as pumped-storage plants without external contributions of water. Then, the energy stored only depends on the pumping and turbine power balance. The PS-HPPs have been modelled as follows:

1. Maximum power generation/consumption: the same rated power has been considered for pumping and turbine power.

$$x_{GHk,t} \leq G_{Hk} \quad \forall k, \forall t \quad (3)$$

$$x_{PHk,t} \leq G_{Hk} \quad \forall k, \forall t \quad (4)$$

where $k$ refers to each of the $H_k$ hydropower plants, $x_{GHk,t}$ and $x_{PHk,t}$ are the instantaneous generation and pumping power of the hydropower plant $H_k$ at the time $t$, respectively, and $G_{Hk}$ is the installed capacity of the hydropower plant $H_k$.

2. Energy balance of the storage system:

$$x_{Sk,t+1} = x_{Sk,t} + x_{PHk,t} \cdot \eta_P - \frac{x_{GHk,t}}{\eta_G} \quad \forall k, \forall t \in [1, T-1] \quad (5)$$

where $x_{Sk,t}$ is the energy stored in the hydropower plant $H_k$ at the time $t$ and $\eta_P$ and $\eta_T$ are the pumping and turbine efficiencies, respectively.

3. Maximum energy storage: defined by the capacity of the upper reservoir of the hydropower plant.

$$x_{Sk,t} \leq S_{Hk} \quad \forall k, \forall t \quad (6)$$

where $S_{Hk}$ is the storage capacity of the hydropower plant $H_k$.

*4.2.4 Solar thermal power plants*

The solar thermal power generation represented in this study is based on parabolic troughs, as it is the main technology used in the solar thermal power plants (SolarPACES, 2021). These power plants are usually equipped with thermal storage systems, which are sized to have the capability to maintain rated power for several hours. Then, the solar thermal generation has been represented including the thermal power absorbed by the parabolic troughs, the thermal storage and the electric generation:

1. Maximum thermal power absorbed by the solar field: the thermal power absorbed by the parabolic troughs highly depends on the solar irradiation and the angle of incidence (Biencinto, González, Valenzuela & Zarza, 2019):

$$x_{TSl,t} \leq i_{Sl,t} \cdot r_l \cdot G_{Sl} \cdot \eta_{O,l} \cdot \eta_{Ef} \cdot K(\theta) \quad \forall l, \forall t \quad (7)$$

where $l$ refers to each of the $S_l$ solar thermal power plants, $x_{TSl,t}$ is the thermal power absorbed by the $S_l$ solar thermal power plant at time $t$, $i_{Sl,t}$ is the solar irradiation in kW/m², $r_l$ is a ratio which relates the solar field surface needed to generate 1 kWe (electric power kW) (see (Biencinto, González, Valenzuela & Zarza, 2019)), $G_{Sl}$ is the rated electrical power of the plant, $\eta_{O,l}$ is the peak optical efficiency, $\eta_{Ef}$ is an efficiency factor which considers other losses, such as thermal, cleanliness or tracking losses, and $K(\theta)$ is a factor obtained from the angle of incidence. $K(\theta)$ can be calculated as (Fernandes & Ferreira, 2014):

$$K(\theta) = 1 - \frac{7 \cdot 10^{-4} \cdot \theta + 36 \cdot 10^{-6} \cdot \theta^2}{cos\theta} \quad (8)$$

where $\theta$ is the angle of incidence, which can be found for a specific location and time in (National Oceanic and Atmospheric Administration (NOAA), 2021).

2. The Maximum electric power is restricted by the rated power:

$$x_{GSl,t} \leq G_{Sl} \quad \forall l, \forall t \qquad (9)$$

where $x_{GSl,t}$ is the electric power generation of the solar thermal power plant $S_l$ at the time $t$.

3. Energy balance of the thermal storage system: the thermal energy stored varies based on the thermal and electric powers as:

$$x_{Sl,t+1} = x_{Sl,t} + x_{TSl,t} - \frac{x_{GSl,t}}{\eta_{th,l}} \quad \forall l, \forall t \in [1, T-1] \qquad (10)$$

where $x_{Sl,t}$ is the energy stored in the solar thermal power plant $S_l$ at the time $t$ and $\eta_{th}$ is the thermoelectric efficiency of the thermal power plant. An ideal storage system has been assumed, so storage losses are not considered.

4. Maximum thermal energy storage: the maximum thermal energy is defined by the capacity of the storage tank:

$$x_{Sl,t} \leq S_{Sl} \quad \forall l, \forall t \qquad (11)$$

where $S_{Sl}$ is the storage capacity of the solar thermal power plant $S_l$.

### 4.2.5 Power system

The power system has been modelled using and aggregated representation, which only considers the total system demand. The grid equations are not included in the model. Then, two restrictions have been implemented:

1. Generation-demand balance: the power generation must meet the total system demand for every time interval.

$$\sum_{i=1}^{I} x_{GCi,t} + \sum_{j=1}^{J} x_{GRj,t} + \sum_{k=1}^{K} x_{GHk,t} + \sum_{l=1}^{L} x_{GSl,t} = D_t + \sum_{k=1}^{K} x_{PHk,t} \qquad (12)$$

where $I$ is the number of conventional power plants, $J$ is the number of PV and wind power plants, $K$ is the number of pumping hydropower plants and $L$ is the number of solar thermal power plants.

2. Minimum contribution of renewable generation during a year:

$$\sum_{t=1}^{T}\left(\sum_{j=1}^{J} x_{GRj,t} + \sum_{l=1}^{L} x_{GSl,t}\right) >= \alpha \sum_{t=1}^{T} D_t \qquad (13)$$

where $\alpha$ is the minimum share of renewable generation expressed in per unit. Pumping hydro generation is not included as renewable generation, as its net energy contribution is null or even negative if the pumping and turbine efficiencies are considered.

## 4.3 Optimization problem

The optimization algorithm provides the generation mix that minimizes the cost for the system. Then, the objective function $f_{obj}(x)$ of the optimization function can be defined as:

$$f_{obj}(x) = OPEX_{Conv}(x) + CAPEX_{Ren}(x) + OPEX_{Ren}(x)$$

$$\text{subject to } h_m(x) = 0, \ m \in [1, M] \ g_n(x) \leq 0, \ m \in [1, N] \qquad (14)$$

where $x$ is the variable vector, $OPEX_{Conv}(x)$ is the operation cost of the conventional power plants, $CAPEX_{Ren}(x)$ is the capital cost of the renewable generation that must be installed, $OPEX_{Ren}(x)$ is the operation cost of the renewable generation, $h_m(x)$ are the equality constraints and $g_n(x)$ are the inequality constraints.

The variable vector $x$ includes the instantaneous generation for every time interval for all the generation types considered, as well as the installed capacity of the renewable generation. Then, the variable vector can be defined as:

$$x = [x_C, C_R, x_R, C_H, x_H, C_S, x_S] \qquad (15)$$

where

$$x_C = [x_{GCi,1}, \ldots, x_{GCi,T}] \quad \forall i \qquad (16)$$

$$C_R = [C_{R1}, \ldots, C_{RJ}] \qquad (17)$$

$$x_R = [x_{GRj,1}, \ldots, x_{GRj,T}] \quad \forall j \qquad (18)$$

$$C_H = [C_{H1}, \ldots, C_{HK}] \qquad (19)$$

$$x_H = [x_{GHk,1}, \ldots, x_{GHk,T}, x_{PHk,1}, \ldots, x_{PHk,T}] \quad \forall k \quad (20)$$

$$C_S = [C_{S1}, \ldots, C_{SL}] \quad (21)$$

$$x_S = [x_{GSl,1}, \ldots, x_{GSl,T}] \quad \forall l \quad (22)$$

The cost functions are defined as:

$$OPEX_{Conv}(x) = \sum_{t=1}^{T} \left( \sum_{i=1}^{I} x_{GCi,t} \cdot OPEX_{Ci} \right) \quad (23)$$

$$CAPEX_{Ren}(x) = \sum_{j=1}^{J} C_{Rj} \cdot CAPEX_{Rj} + \sum_{k=1}^{k} C_{Hk} \cdot CAPEX_{Hk} + \sum_{l=1}^{L} C_{Sl} \cdot CAPEX_{Sl} \quad (24)$$

$$OPEX_{Ren}(x) = \sum_{t=1}^{T} \left( \sum_{j=1}^{J} x_{Rj} \cdot OPEX_{Rj} + \sum_{k=1}^{K} x_{Hk} \cdot OPEX_{Hk} + \sum_{l=1}^{L} s_{Sl} \cdot OPEX_{Sl} \right) \quad (25)$$

The equality and inequality constraints, $h_m(x)$ and $g_n(x)$, are based on the models of the different generation types described in Section 4.2.

## 5. APPLICATION OF THE METHODOLOGY TO THE SELECTED SCENARIOS

The previous methodology has been applied to some of the realistic scenarios shown in Section 3. These scenarios have been studied in the context of the POSYTYF project (POSYTYF, 2020). Two cases have been selected in order to validate the methodology and exemplify the sizing of such scenarios:

- Scenario 1: Type I – Isolated: island
- Scenario 3: Type II – Synchronously interconnected (AC): northern Europe

### 5.1 Scenario 1: Type I – Island

Tenerife has been chosen as the reference location to obtain the system demand and solar and wind resources availability. The hourly power demand in 2019 varied from 300 to 550 MW approximately (Red Eléctrica de España (REE), 2021). It has been assumed that a 600-MW coal-fired generation is already installed. Regarding RES, two possible cases have been considered for this scenario:

- Case 1 - without storage: only PV and wind are considered.
- Case 2 - with storage: in addition to PV and wind RES generation technologies, solar thermal generation with storage has been included in the system.

*5.1.1 Case 1 – without storage*

First, the optimization has been executed to obtain the RES installed capacity for a 74% share of renewables ($\alpha$), which is the target established by the Spanish government for 2030 (Spanish government, 2020). To meet the previous requirement, the optimization algorithm has determined that wind and PV generation technologies should provide 43% and 31% of the total demand, respectively (see Figure 16). Furthermore, the overall installed capacity of wind and PV should be 785 MW and 596 MW respectively (Figure 16).

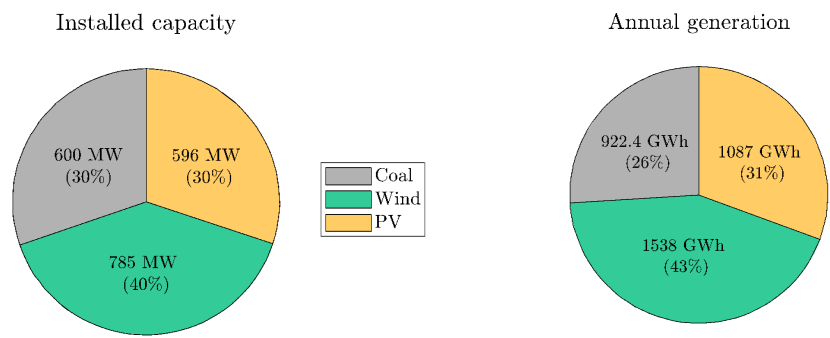

Figure 16. Installed capacity and annual generation mix for Case 1 and $\alpha = 74\%$.

Figure 17 shows the generation mix for 2019 on a daily and monthly basis. It can be noticed that the PV generation pattern is very similar during the whole year. On the other hand, wind energy generation varies considerably and coal-fired generation compensates the power variations. Regarding per month generation, July and August are the months with higher contribution of wind energy, and January and October for coal-based generation.

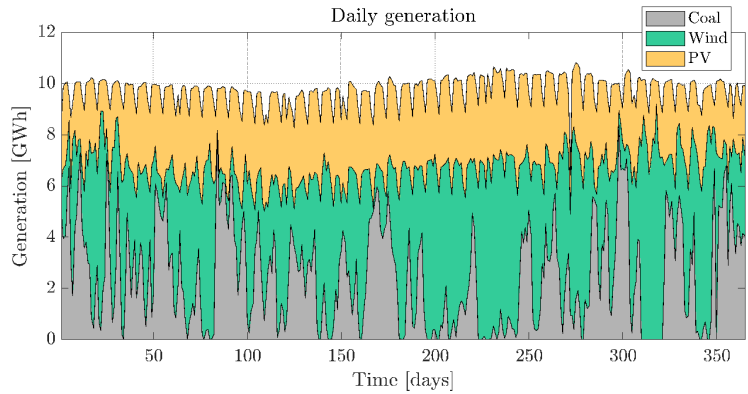

(a)

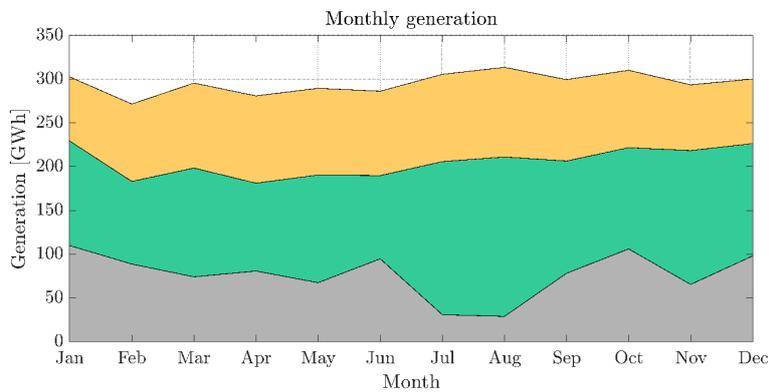

(b)

Figure 17. Generation mix for Case 1 in 2019: (a) daily; (b) monthly.

Additionally, the effect of the minimum share of renewables ($\alpha$) on RES installed capacity has also been analyzed in Figure 18. It can be observed the exponential curve, requiring an extremely high installed capacity for $\alpha$ close to 100%. This is caused by the lack of solar and wind resources at some time intervals, *i.e.*, low wind at night, forcing the algorithm to oversize them. For $\alpha = 100\%$, no solution is found as neither PV nor wind generation can supply the demand.

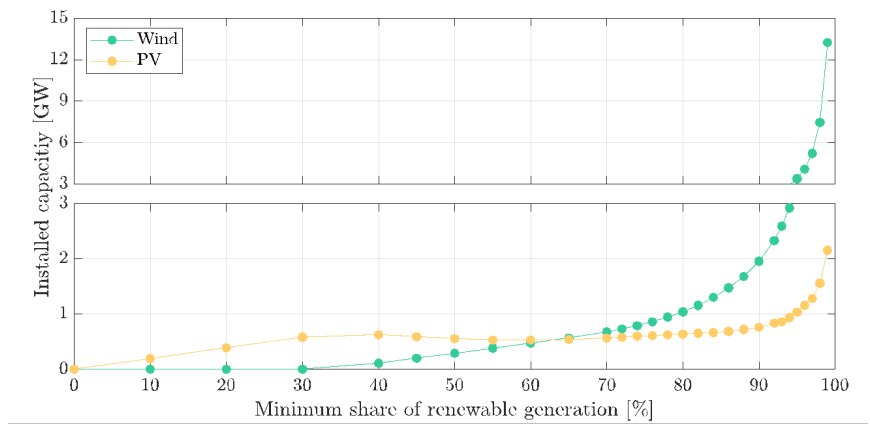

Figure 18. Installed capacity of wind and PV generation for Case 1

Based on the results obtained and the preliminary layout from Section 3 (Figure 11), the final scenario is depicted in Figure 19. The scenario contains a total amount of 7 transmission busses and no distribution. The voltage levels of the transmission lines could be, for instance, in the range of 66-220 kV. The obtained capacity for wind and PV is distributed in the different available busses, and two conventional generation units are considered. Different portions of the demand are distributed among the busses. The distances are relatively short, as this scenario represents a relatively small island.

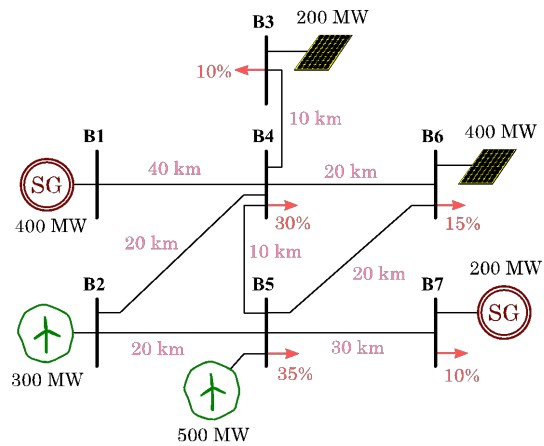

Figure 19. Scenario 1: Island without storage capability

### 5.1.2 Case 2 – with storage

To overcome the previous oversizing issue, storage elements can be introduced into the system. In this case, solar thermal power generation has been included in the optimization problem. A thermal storage system equivalent to 7.5 h of rated electric power has been assumed. The actual storage capacity will depend on the rated power of the solar thermal generation, which is also a variable of the optimization problem.

Then, a similar analysis has been carried out to study the renewable generation that must be installed for a different $\alpha$, resulting in Figure 20. Due to the higher cost of solar thermal generation, the algorithm does not include it in the solution until $\alpha \geq 80\%$. Then, for $\alpha$ lower than 80 %, Figure 20 is identical to Figure 18. For $\alpha$ above this value, the introduction of solar thermal generation avoids the oversizing of wind and PV generation. However, for $\alpha = 100\%$, the optimization provides a non-realistic solution. More than 30 GW of wind energy is required to supply a system with a peak demand of around 550

MW (this singularity is not represented in Figure 20). For $\alpha = 99.9\%$, the obtained solution is still acceptable, although the required renewable capacity is considerably higher than those obtained for 99%. Figure 21 shows the generation cost comparison between Cases 1 and 2. The introduction of storage into the system allows a considerable cost reduction for $\alpha$ higher than 90%.

A specific case when $\alpha = 90\%$ has been selected to exemplify the system's performance when the solar thermal power plant is included. Figure 22 shows the installed capacity and the annual generation for every technology. It can be observed how renewables can supply 90% of the energy while their contribution to the installed capacity is only 75%. This is possible thanks to the storage system of the solar thermal power plant.

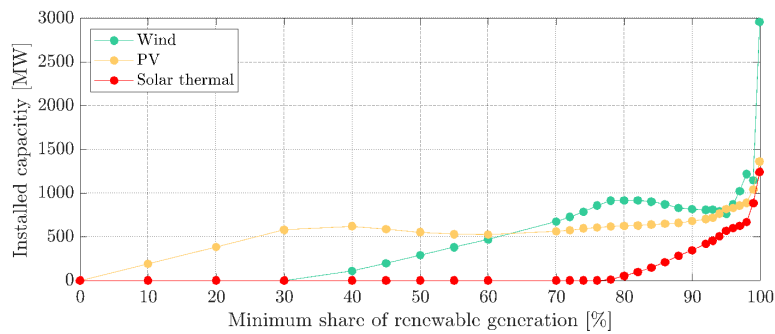

Figure 20. Installed capacity of wind, PV and solar thermal generation for Case 2.

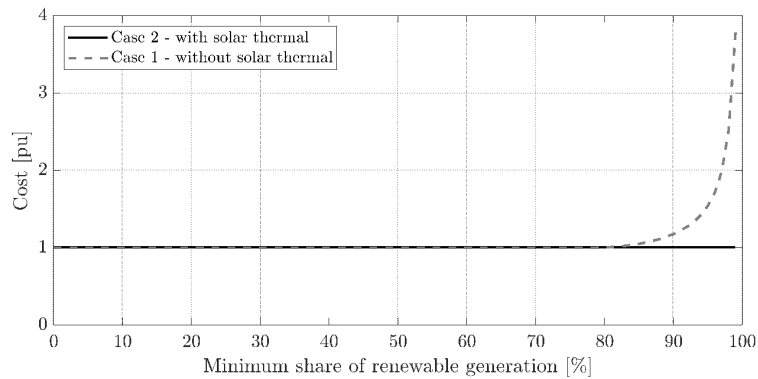

Figure 21. Total generation cost considering Case 2 as the base case.

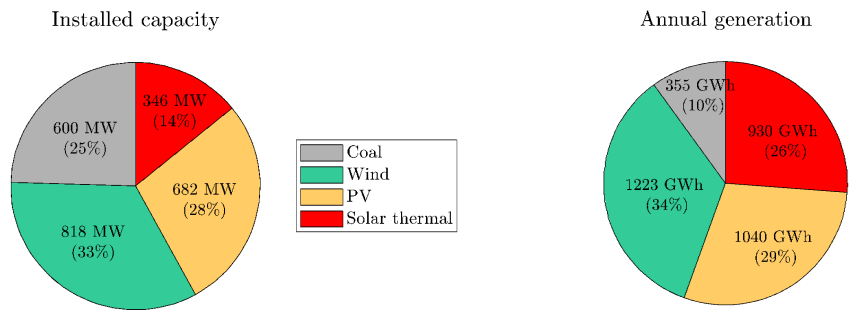

Figure 22. Installed capacity and annual generation mix for Case 2 and $\alpha = 90\%$.

Additional specific results about daily and monthly generation are shown in Figure 23. PV and solar thermal generation contribution are higher in summer, while coal has to compensate for the lack of solar resources in winter.

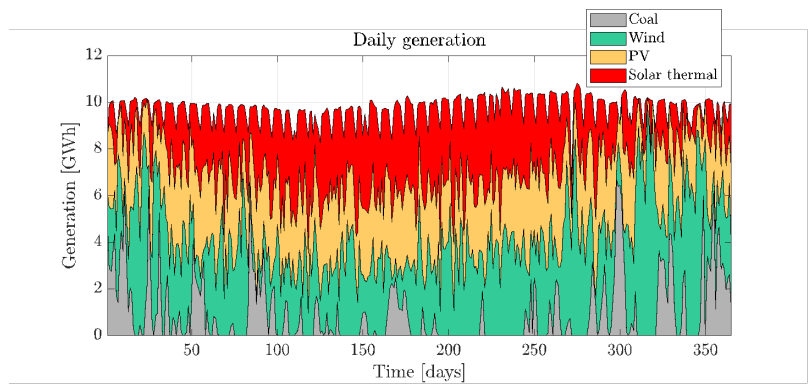

(a)

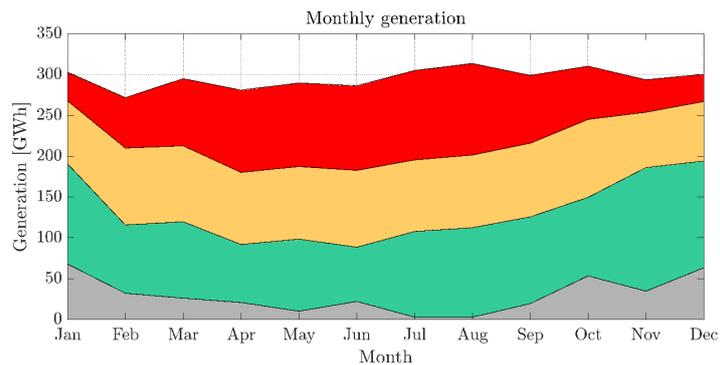

(b)

Figure 23. Generation mix for Case 2 in 2019: (a) daily; (b) monthly.

The same grid configuration is considered in this second case but including solar thermal (Figure 24). The conventional, wind and PV obtained capacities are very similar to the case without solar thermal, and thus the powers shown are the same as in the previous case. The additional solar thermal capacity is distributed equally in busses 4 and 6, in the same area where the PV is located.

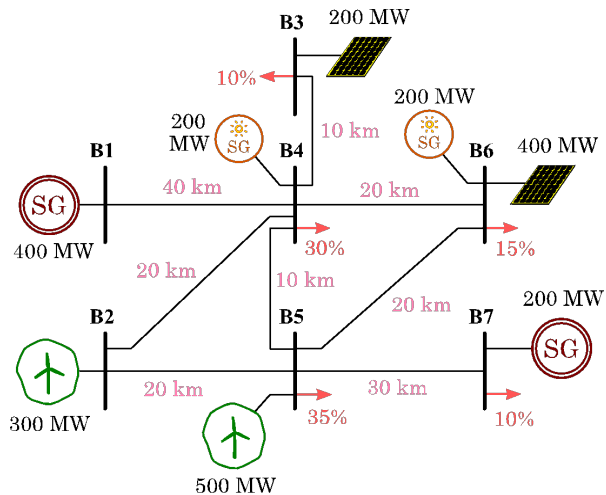

Figure 24. Scenario 2: Island without storage capability – Final proposed layout.

*5.1.3 Scenario 3: Type II – Synchronously interconnected (AC): northern Europe*

Scenario 3 corresponds to an AC interconnected system (Type II) located in northern Europe. This scenario has been specifically located in the Netherlands to extract the system demand and the solar and wind resources. The hourly consumption data of the Netherlands obtained from (Open Power System Data, 2021) has been scaled down to have a maximum instantaneous demand of 5 GW, leading to a minimum demand of 2.37 GW.

The generation technologies included in this scenario are coal, wind, PV and PS-HPP. Three cases have been considered to analyze the effect of water storage in the system:

- Case 1: no storage.
- Case 2: PS-HPP generation with an installed capacity equal to the 10% of the maximum demand.
- Case 3: PS-HPP generation with an installed capacity equal to the 20% of the maximum demand.

For Cases 2 and 3, PS-HPP generation is assumed to be already installed, so the capital cost is not considered.

Figure 25 shows the wind and PV capacity required based on the minimum renewable share. It is observed that PS-HPP generation presence can help to reduce the amount of renewable generation considerably. For $\alpha = 90\%$, the wind capacity obtained for Case 1 is around 24 GW, while it is reduced to 19 GW for Case 2 and 15 GW for Case 3. So, a reduction of almost 10 GW of wind power can be achieved only by 1 GW of PS-HPP generation.

This results in a reduction of the generation costs of the system., shown in Figure 26. The costs have been normalized considering Case 3 as the base case. It can be observed that Case 3 provides a cost around 50% and 20% lower than Case 1 and Case 2 when $\alpha = 90\%$. Higher cost reductions are achieved for higher values of $\alpha$, as PS-HPP generation avoids the installation of new wind and PV generation. However, it must be noted that the capital cost of PS-HPP generation has not been considered. In that case, the cost reduction obtained would be lower.

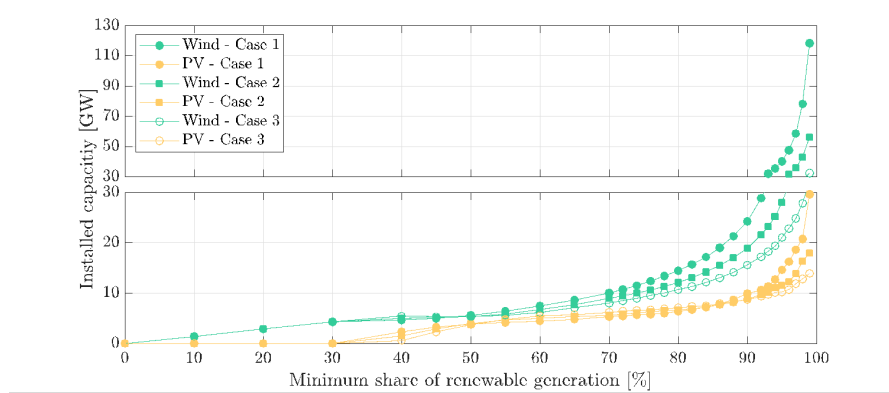

Figure 25. Installed capacity of wind and PV.

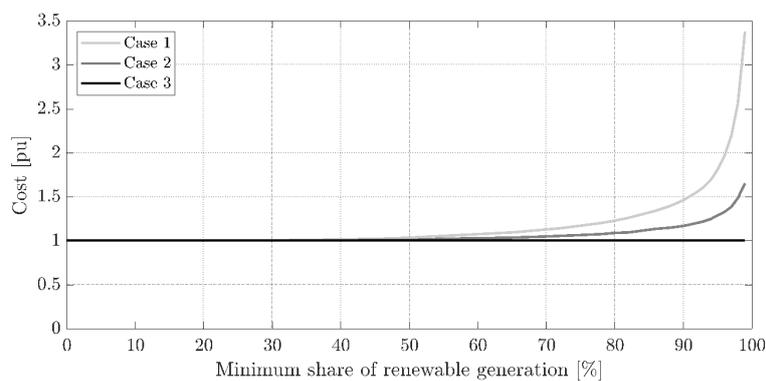

Figure 26. Total generation cost considering Case 3 as the base case.

The generation mix respect to $\alpha$ is depicted in Figure 27 for the three cases. Coal, wind and PV generation share the demand, while the hydro generation is always above 100%, as the net energy contribution of PS-HPP is null. The use of hydro generation rises when $\alpha$ is increased, allowing storing energy and saving the installation of wind or PV. In Cases 2 and 3, the coal generation is different from zero when $\alpha$ is set to 100%. The restriction in (13) ensures that wind and PV generation is equal to the system demand. When PS-HPP is included, this demand is increased due to the pumping consumption, which is partially supplied by coal.

Further analysis has been carried out when $\alpha = 74\%$. Figure 28 shows the installed capacity and annual generation for all cases. The installation of PS-HPP allows supplying nearly the same amount of renewable generation reducing the installed capacity of wind and PV for Cases 2 and 3.

Specific daily, monthly and hourly results of Case 3 are shown in Figure 29. In this case, the PV generation varies throughout the year, as Scenario 3 is located in a higher latitude than Scenario 1. Figure 29.c shows how the pumping is used when there is a high renewable generation, helping to reduce the generation cost.

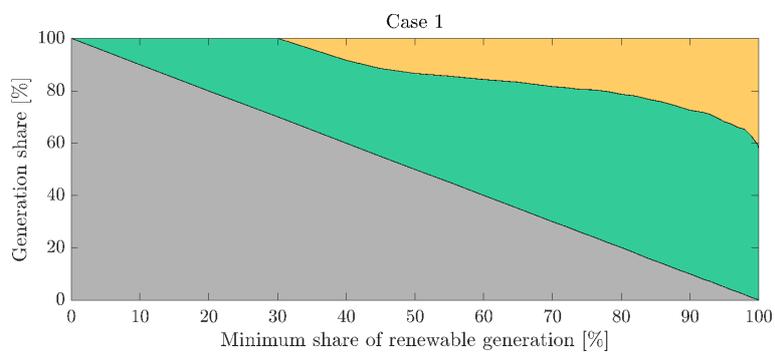

(a) Case 1

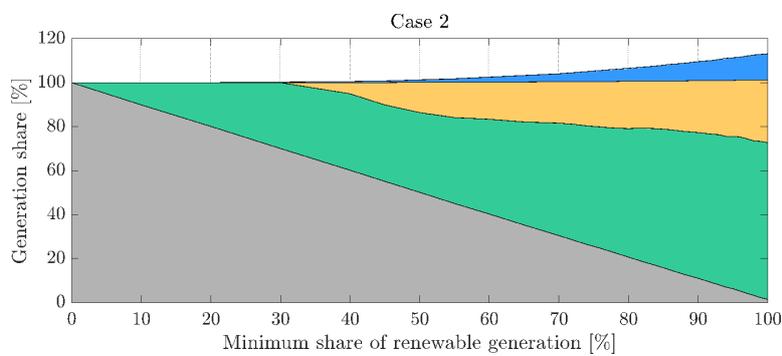

(b) Case 2

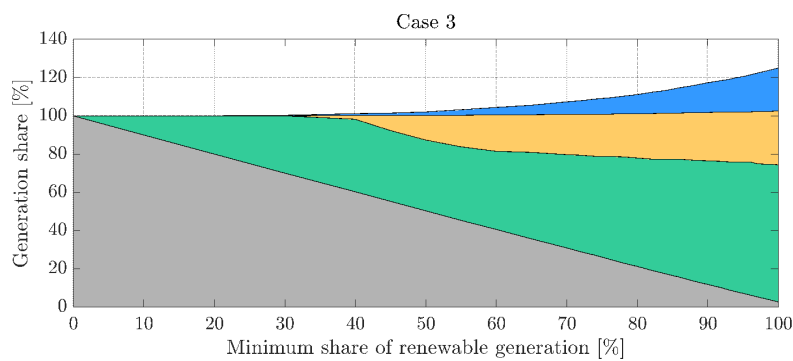

(c) Case 3

Figure 27. Generation share based on the minimum renewable required.

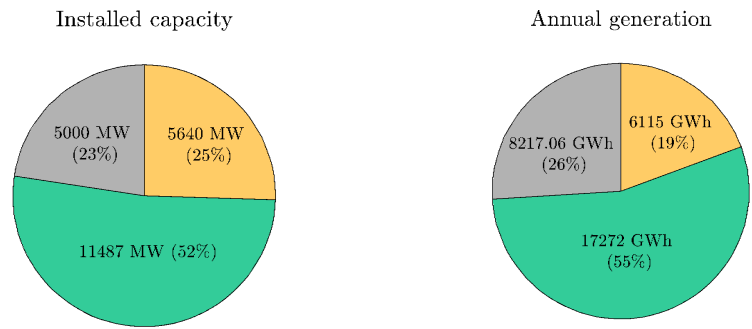

(a) Case 1

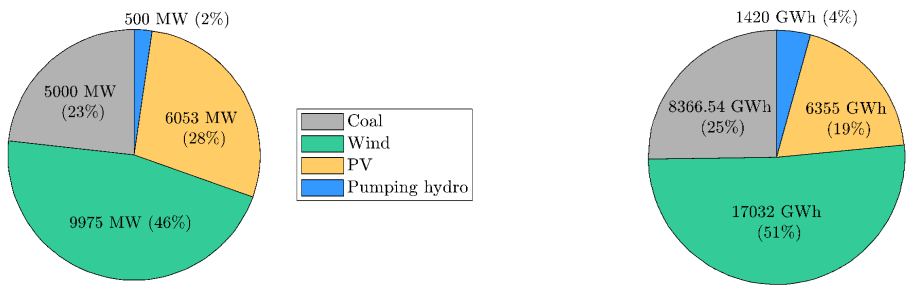

(b) Case 2

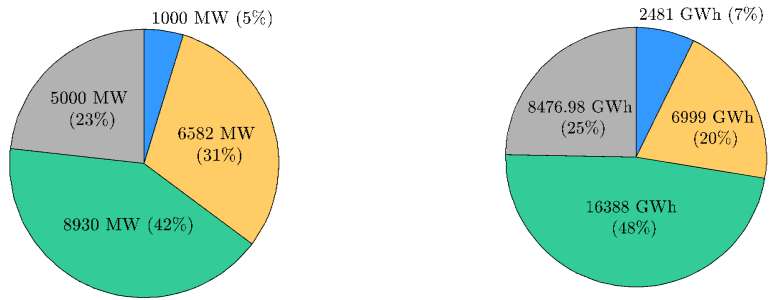

(c) Case 3

Figure 28. Installed capacity and annual generation mix for $\alpha = 74\%$.

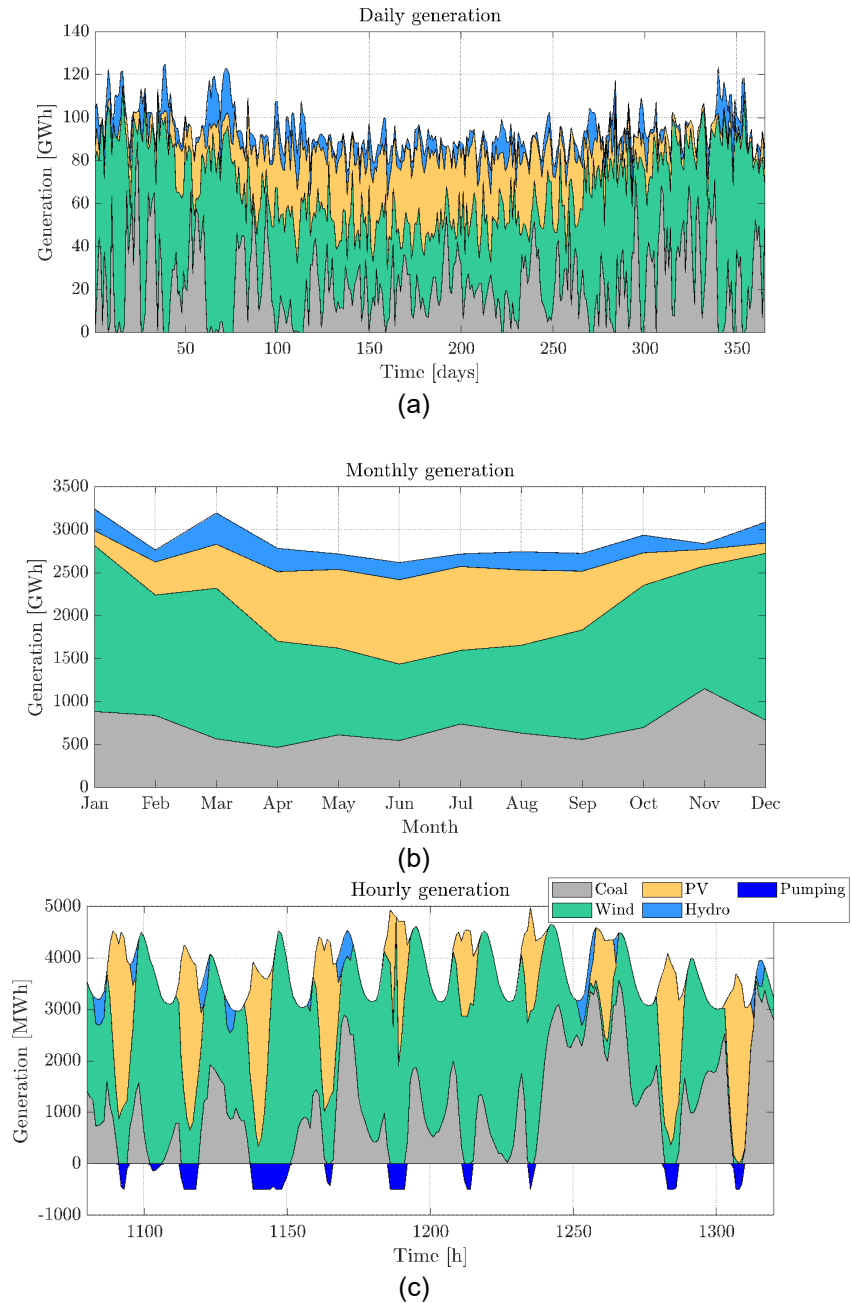

Figure 29. Daily, monthly and hourly generation for Case 3 and $\alpha = 74\%$.

Based on the results obtained and the preliminary layout from Section 3 (Figure 13), the final scenario is depicted in Figure 30. The scenario contains a total amount of 8 transmission busses and 5 distribution busses. The large size of the different power plants does not represent a single power plant but an aggregated equivalent of several ones. The voltage levels of the transmission lines could be, for instance, in the range of 220-400 kV, whereas in the distribution case, it could be 20-30 kV. An appropriate amount of offshore wind (both DC-interconnected and AC-interconnected) is considered, as this scenario is inspired in the north of Europe. Also, onshore wind, PV and PS-HPP are considered, and a portion of conventional generation.

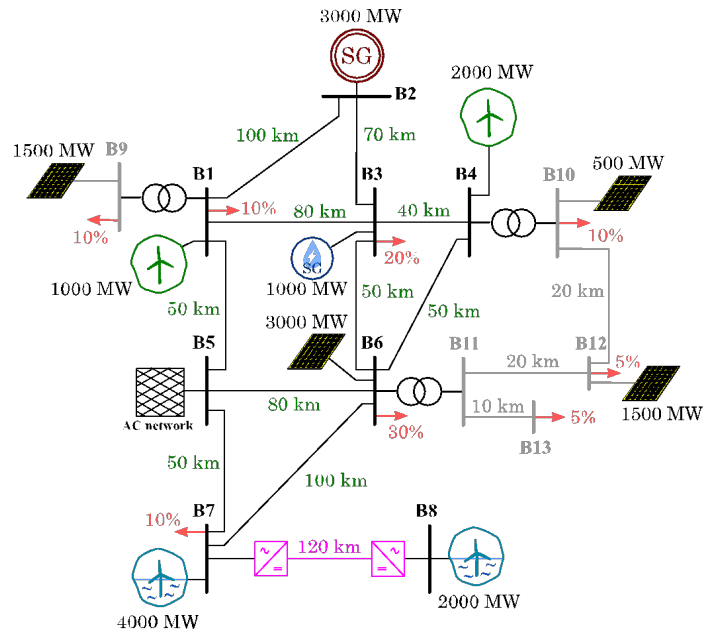

Figure 30. Scenario 3 – Final proposed layout.

**Conclusion**

This overview has presented different possible scenarios that can be used for the analysis of large integration of RES in Europe. The design of the scenarios has been done considering specific weather conditions and renewable resources of specific regions, and an optimization-based methodology has been used to quantify the amount of renewable generation capacity needed. Different renewable energy technologies have been considered, in order to meet specific requirements of grid integration of renewables at different horizons of time, up to 100 % in the most futuristic case.

The optimization algorithm was exampled in three scenarios, considering and not considering storage. It has been shown that some technologies can provide the renewable backbone (solar PV and wind) but they lack the flexibility needed to achieve a very high share in the energy mix. Other technologies become important to cover the last range of integration (for instance, solar thermal and pumped hydro), as they provide a high flexibility, which is crucial for high share, but they are expensive for low share. Otherwise, if these technologies are excluded, the required installed capacity if, for instance, only wind and PV are considered, might rise substantially for high constraints of renewable share.

The proposed scenarios can be considered realistic in the sense that they are inspired by real data and real locations, but they are not detailed in the sense of power system operation, rather serving as a starting point to future studies. As the current power system still contains a large amount of conventional thermal power plants, the current network configuration and the presence of renewable power plants might be subject to important changes over the next years and decades.

Based on the optimization results applied to the analysed scenarios, the future European targets that consider a generation mix composed mainly by renewable generation will invariably require the participation of storage technologies in the grid, to reduce the ratio between installed RES capacity and maximum demand and increase the system flexibility.

**Funding Information**

This work was implemented in POSYTYF project, supported by the European Union's Horizon 2020 research and innovation programme under grant agreement No 883985. The work of Oriol Gomis-Bellmunt is supported by the ICREA Academia program. Eduardo Prieto-Araujo and Marc Cheah-Mañe are Serra Hunter lecturers.